\documentclass[%
 reprint,
superscriptaddress,
amsmath,amssymb,
aps
twocolumn
]{revtex4-2}

\usepackage{graphicx}
\usepackage{dcolumn}
\usepackage{bm}

\usepackage[version=4]{mhchem}
\usepackage{color}
\usepackage{tabularx}

\begin{document}

\title[]{Tuning superconductivity in nanosecond laser annealed boron doped $Si_{1-x}Ge_{x}$ epilayers}

\author{S. Nath}
\author{I. Turan}
\author{L. Desvignes}
\author{L. Largeau}
\author{O. Mauguin}
\affiliation{Uni. Paris-Saclay, CNRS, Centre de Nanosciences et de Nanotechnologies, 91120, Palaiseau, France}
\author{M. T\'unica}
\author{M. Amato}
\affiliation{Uni. Paris-Saclay, CNRS, Laboratoire de Physique des Solides, 91405 Orsay, France}
\author{C. Renard}
\author{G. Hallais}
\author{D. Débarre}
\affiliation{Uni. Paris-Saclay, CNRS, Centre de Nanosciences et de Nanotechnologies, 91120, Palaiseau, France}
\author{F. Chiodi}
\email[ Corresponding author: ]{francesca.chiodi@c2n.upsaclay.fr}
\affiliation{Uni. Paris-Saclay, CNRS, Centre de Nanosciences et de Nanotechnologies, 91120, Palaiseau, France}

\date\today

\begin{abstract}
	\noindent
Superconductivity in ultra-doped $Si_{1-x}Ge_{x}:B$ epilayers is demonstrated by nanosecond laser doping, which allows introducing substitutional B concentrations well above the solubility limit and up to $7\,at.\%$.  A Ge fraction $x$ ranging from 0 to 0.21 is incorporated in $Si:B$ : 1) through a precursor gas by Gas Immersion Laser Doping; 2) by ion implantation, followed by nanosecond laser annealing; 3) by UHV-CVD growth of a thin Ge layer, followed by nanosecond laser annealing. The 30 nm and 80 nm thick $Si_{1-x}Ge_{x}:B$ epilayers display superconducting critical temperatures $T_c$ tuned by B and Ge between 0 and 0.6 K. Within BCS weak-coupling theory, $T_c$ evolves exponentially with both the density of states and the electron-phonon potential. While B doping affects both, through the increase of the carrier density and the tensile strain, Ge incorporation allows addressing independently the lattice deformation influence on superconductivity. To estimate the lattice parameter modulation with B and Ge, Vegard's law is validated for the ternary $SiGeB$ bulk alloy by Density Functional Theory calculations. Its validity is furthermore confirmed experimentally by X-Ray Diffraction. We highlight a global linear dependence of $T_c$ vs. lattice parameter, common for both $Si:B$ and $Si_{1-x}Ge_{x}:B$, with $\delta T_c/T_c \sim 50\,\%$ for $\delta a/a \sim 1\,\%$.
\end{abstract}

\maketitle

\section{Introduction}
$SiGe$ is a key material for micro-electronics. The possibility to combine classical SiGe technology with quantum circuits is appealing to exploit the large-scale integration and reproducibility associated with CMOS devices  \cite{Ruffino2022}. Hole spin qubits have been developed in Ge/SiGe quantum dots \cite{Hendrickx2021} and SiGe nanowires~\cite{Hu2012,Amato2014}, taking advantage from the control on the environment and low nuclear spin possible in group IV materials. Furthermore, Ge and SiGe have been incorporated in Josephson field effect transistors ~\cite{Vigneau2019,Amato2014}, hosted transmon qubits, and their microwave losses have been investigated \cite{Sandberg2021}.
The possibility of inducing superconductivity directly in thin SiGe layers might furthermore provide an advantage in coupling SiGe-based classical electronics to superconducting quantum circuits.\\
It has been shown that Silicon displays a superconducting phase when ultra-doped with B \cite{Bustarret2006, Marcenat2010, Grockowiak2013, Chiodi2021}. An extreme boron doping concentration is required to trigger superconductivity in $Si:B$, more than three times the solubility limit. This concentration, impossible to reach using conventional micro-electronic techniques, was obtained using Gas Immersion Laser Doping (GILD), an out-of-equilibrium technique combining chemisorbtion of a precursor gas in a Ultra-High-Vacuum environment, and nanosecond laser annealing \cite{Carey1989, Kerrien2002, Chiodi2021}.
In this paper, we employ this technique to ultra-dope with B thin $SiGe$ layers, demonstrating the realisation of a superconducting phase.\\
In addition to the intrinsic interest of $SiGe$ superconductivity, the investigation of the evolution of the superconducting critical temperature $T_c$ with both B and Ge doping allows to better understand what triggers superconductivity in Si and SiGe. Indeed, BCS theory in the weak coupling limit expects $T_c$ to exponentially increase with the electron-phonon coupling $\lambda = N(E_F) V_{e-ph}$, the product of the electron-phonon potential $V_{e-ph}$ and the density of states at Fermi energy $N(E_F)$. B doping modifies both $N(E_F)$ and $V_{e-ph}$: $N(E_F)$ is related to $n_B$, as evident in the frame of the free electron model where $E_F \propto (3\pi^2 n_B)^{2/3}$. But, due to the smaller B size, an important lattice deformation up to $\delta a/a = -3.5\,\%$ is induced at the same time \cite{Marcenat2010}, affecting the phonon frequencies involved in Cooper pairing. The incorporation of Ge makes it possible to modify, solely and independently, the lattice deformation, allowing to elucidate the relevant parameters that govern superconductivity.
\begin{table*}{}
    \centering
    \setlength{\tabcolsep}{0.5em}
    \bgroup
    \def\arraystretch{1.3}
    \begin{tabular}{| c || c | c | c | c | c |}
    \hline
      Ge incorporation   & Thickness (nm) & Laser shots Ge  & $C_{Ge}$ (at.$\%$) & Laser shots B & $C_{B}$ (at.$\%$)  \\
      \hline
      GILD - 5 & 30 & 5 & 0.27 & 30-400 & 2.3 - 10.6 \\
      \hline
      GILD - 15  & 30 & 15 & 0.8 & 30-400 & 2.3 - 10.6 \\
      \hline
      GILD - 200  & 30 & 200 & 10.7 & 160-475 & 8 - 11.2 \\
      \hline
      GILD - Ge & 30 & 5-400 & 0.27-21.3 & 220 & 9.2 \\
      \hline
      Ge CVD  & 30 & - & 17.1  & 30-400 & 2.3 - 10.6 \\
      \hline
      Ge implanted ($10^{15}\,cm^{-2}$)  & 75 & - & 0.27 & 3 - 258 + B impl. ($5 \times 10^{15}\,cm^{-2}$) & 1.5 - 10.8 \\
      \hline
      Reference $SiB$  & 30 & - & -  & 50-400 & 3.6 - 10.6 \\
      \hline
    \end{tabular}
    \egroup
    \caption{Details on the $SiB$ and $SiGeB$ sample series investigated in this work: Ge incorporation method, thickness, number of nanosecond laser annealing repetitions, total B and Ge concentrations $C_B$ and $C_{Ge}$ in atomic $\%$.   }
    \label{Table}
\end{table*}
\section{Ultra doped $Si_{1-x}Ge_{x}:B$ }
\subsection{Gas Immersion Laser Doping}
To attain the extremely high doping levels necessary to overcome the superconductivity threshold, we employ fast liquid phase epitaxy, characterized by recrystallization times of a few tens of nanoseconds (see Methods).
A puff of the precursor gas ($BCl_3$ or $GeCl_4$) is injected onto the substrate surface, saturating the chemisorption sites, so that the supply of incorporated atoms is constant and self-limited. A pulse of excimer XeCl laser ($\lambda$ = 308 nm, pulse duration 25 ns) melts the substrate, and the chemisorbed atoms diffuse in the liquid. At the end of the laser pulse, an epitaxial out-of-equilibrium recrystallization takes place from the substrate at a speed of $\sim 4\,$m/s \cite{Wood1984}, achieving concentrations larger than the solubility limit ($\sim 1 \,at.\%$ for B in Si).  The laser energy density, tuned with an attenuator, controls the melted thickness in the 5-500 nm range. A flat, straight, and sharp (few nm thick) interface is created between the ultra-doped layer and the substrate. In order to control the amount of B and Ge incorporated, the entire chemisorption-melting-crystallization process is repeated the desired number of times (number of laser shots N). The total B (Ge) concentration $C_B$ ($C_{Ge}$) is proportional to the number of GILD process repetitions $N$ \cite{Bhaduri2012, Bonnet2019, Desvignes2023} and is calibrated by integrating the SIMS (Secondary Ion Mass Spectrometry) concentration profiles to calculate the average concentration in the layer.

\subsection{Ge incorporation}
We explore the low temperature electrical characteristics of thin $Si_{1-x}Ge_{x}:B$ films, ultra-doped in boron (B) by Gas Immersion Laser Doping. Three different methods are used to incorporate the Ge and control its amount: 1) Gas Immersion Laser Doping with $GeCl_4$ as a precursor gas; 2) Implantation of Ge and B, followed by nanosecond laser annealing; 3) Growth by UHV-CVD of a thin Ge layer, followed by nanosecond laser annealing.\\
N-type Si substrates of resistivity $50\,\Omega cm$ are used for all sample series. The substrates are introduced in the UHV chamber after an acetone and ultrasounds cleaning to remove organic surface contamination, and 1 minute Buffered Hydrofluoric acid etch to remove the native silicon oxide.

\textit{1) Gas Immersion Laser Doping of Ge}\\
Three samples series are realized with varying B content and a fixed Ge concentration, and one with a varying Ge content and fixed B concentration, by controlling the respective B and Ge number of GILD process repetitions (see summary table \ref{Table}). The layers doped thickness is $t=30\,$nm, corresponding to an annealing time of $30\,$ns on undoped Si by a laser energy density at the sample level of $1000\,$mJ/cm$^2$. 
B atoms have a high diffusion coefficient in the liquid Si ($D\sim 10^{-4}$cm$^2$/s) and a segregation coefficient near 1 at the high crystallisation speeds attained ($k=0.95$) \cite{Wood1984-5}, insuring a homogeneous B distribution within the recrystallised layer even for the longer annealing processes ($\sim 15\,\mu$s for 500 cycles of 30 ns). Thus, a homogeneous distribution is expected even when the B is further submitted to the subsequent process time of the Ge incorporation.  In contrast, a graded Ge profile is expected, evolving toward the surface and depleting the bottom of the layer, an effect of the smaller segregation coefficient ($k\sim 0.6-0.8$) \cite{Fossard2008}. \\
\textit{2) Nanosecond Laser Annealing of implanted Ge and B} \\
In one sample series, a Ge dose of $10^{15}\,cm^{-2}$ is introduced before the nanosecond laser annealing through an implantation step.
The following GILD processes serve both the purposes of B doping and Ge incorporation in substitutional sites. This series is thicker than the rest of this work ($t=75\,$nm) to ensure that the implanted Ge atoms, including the implantation queue, are within the melted depth. A small implanted B dose ($5 \times 10^{15}\,cm^{-2}$, equivalent to 42 laser shots, i.e. to 1.25 at.$\%$) is also present in the sample and is activated alongside the Ge. B doping is then completed with GILD.\\
\textit{3) Nanosecond Laser Annealing of a thin CVD Ge layer}\\
The epitaxial growth of Ge on Si is carried out in an UHV-CVD system with a base pressure of $10^{-10}$ mbar. Pure $SiH_4$ and $GeH_4$ diluted at 10$\%$ in $H_2$ are used as gas sources. The growth time is settled at 20 min in order to achieve 6$\,$nm of Ge \cite{Halbwax2005,Hallais2023a} (see Methods for further details). During the GILD step to incorporate B, the laser energy is kept initially low and gradually increased over the last 10 process repetitions, in order to limit the time of Ge diffusion towards the surface and achieve a Ge profile as homogeneous as possible.
\section{Superconductivity in $Si_{1-x}Ge_{x}:B$ } 
\subsection{Low temperature measurements} At the end of the B and Ge incorporation, Ti(15 nm)/Au(150 nm) metallic contacts for 4-points measurements are deposited over the doped layers by laser lithography and e-beam evaporation.
The resistance $R$ of a region 150$\,\mu$m wide and 300$\,\mu$m long is extracted from dc V(I) measurements of averaged positive and negative bias current ($I_{dc}$=10 to 50 nA).  $R$ is recorded as a function of temperature (300 K to 0.05 K) and perpendicular magnetic field (0 to 55 mT) in an Adiabatic Demagnetisation Refrigerator setup. After the demagnetisation and relative cool down, the system is left to evolve during the slow ($\sim$1.5 hours) temperature increase, giving a precise evaluation of the superconducting transition temperature.  For the transitions in presence of a magnetic field B, the demagnetisation is stopped at B (instead of decreasing to zero field as in usual demagnetisation cycles), so that the transition is recorded in presence of a constant magnetic field. \\
In order to access the hole carrier density, Hall bars are realized in a separate reference $SiB$ sample series (see Methods). 
\begin{figure}[t]
		\centering
		\includegraphics[width=\columnwidth]{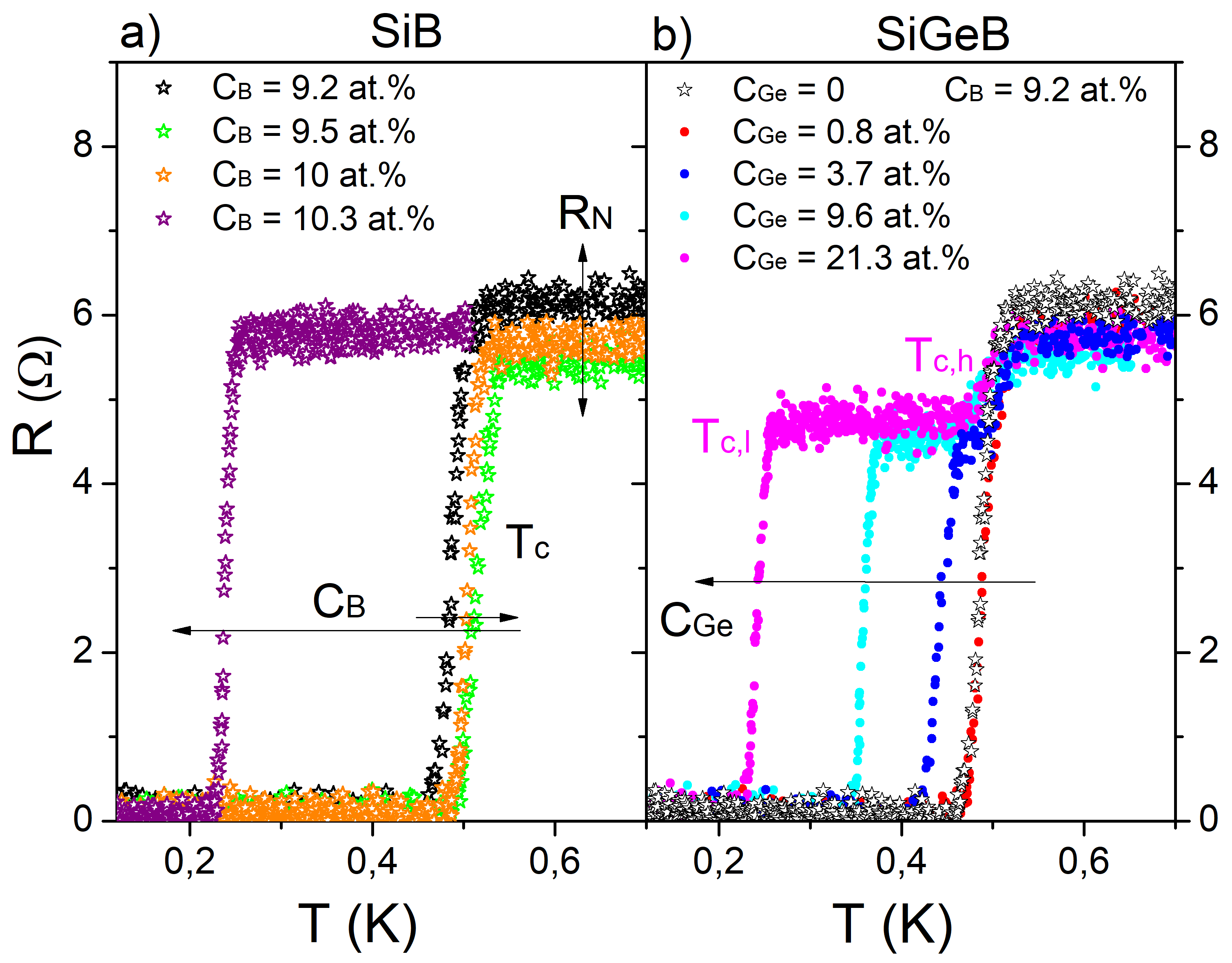}
		\caption{\label{fig:RT_B} Typical resistance $R$ vs. temperature $T$ superconducting transitions for a) the reference $SiB$ GILD samples series, containing no Ge and a variable B concentration $C_{B}=9.2 - 10.3\,$at.$\%$; b)  $SiGeB$ layers from the GILD-Ge sample series, with constant $C_B = 9.2\,$at.$\%$ and varying Ge concentration $C_{Ge}=0.8 - 21.3\,$at.$\%$. $T_{c,h}$ and $T_{c,l}$ indicate respectively the high and low temperature transition for $SiGeB$ at high doping.
  }
	\end{figure}
\subsection{Superconductivity evolution with B in $Si:B$}
\begin{figure}[t]
		\centering
		\includegraphics[width=\columnwidth]{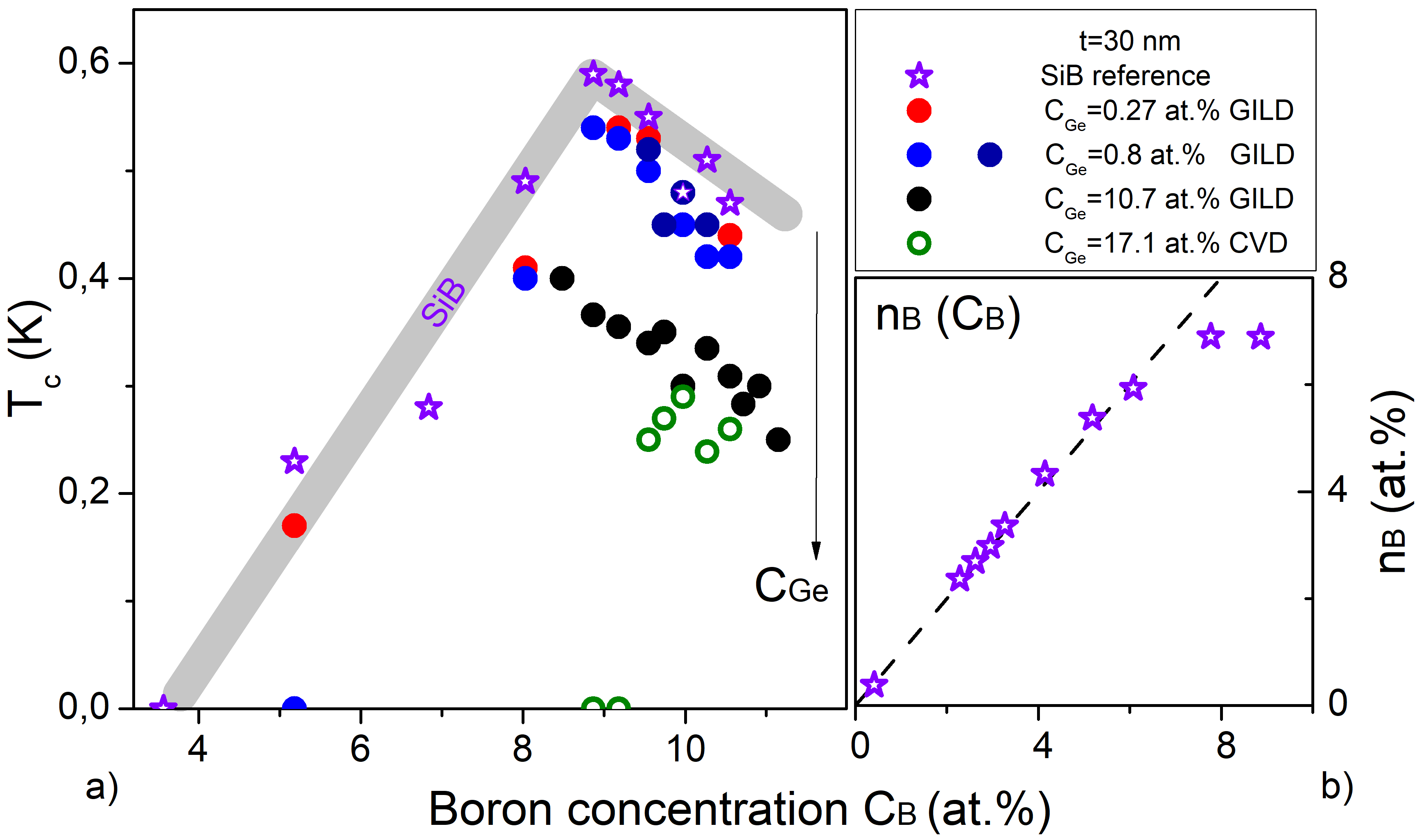}
		\caption{\label{fig:Tc_B} a) $SiB$ and $SiGeB$ superconducting critical temperature $T_c$ at mid-transition vs. total B concentration $C_B$ for sample series GILD-5, GILD-15, GILD-200, Ge CVD and Reference $SiB$. All series are 30$\,$nm thick. $SiGeB$ samples present two transitions at the highest $C_{Ge}$. In this case, the one leading to the zero resistance state, $T_{c,l}$, is plotted.  b) Hole carrier density $n_B$ (substitutional B concentration) vs. total B concentration $C_B$ extracted from Hall measured in a dedicated $SiB$ sample series with t=30 nm. $C_B$ is measured from SIMS (Secondary Ion Mass Spectrometry) concentration profiles over the layer thickness $t$: $C_B = \frac{\int{C_{SIMS}\, dt}}{t}$. The gray line is a guide to the eye for $SiB$ (no Ge) evolution.
  }
	\end{figure}

Fig. \ref{fig:RT_B}-a shows typical $R(T)$ superconducting transitions for the reference $SiB$ GILD samples series, containing no Ge, with varying total B concentration $C_B$. The $R(T)$ curves show a single, relatively sharp transition, of width $\Delta T \sim 0.08\,$K$\sim 16 \%$. We observe that the superconducting critical temperature $T_c$ and the normal state resistance $R_N$ evolve with $C_B$, the total amount of B incorporated. 
Varying $C_B$ results in a modification of the hole carrier density $n_B$. For $C_B < 6 \,at.\%$,  all B atoms are substitutional, providing a hole carrier, and we obtain $100\%$ activation, with $n_B = C_B$ (Fig. \ref{fig:Tc_B}-b). For $C_B > 6 \,at.\%$, the activation progressively lowers with the gradual increase of inactive B complexes, and $n_B$ increase slows down. Finally, at $C_B > 9.5\,$at.$\%$, $n_B$ saturates, as a result of the formation of B aggregates \cite{Desvignes2023, Hallais2023}. As a consequence, $R_N$ initially decreases with $C_B$ in the full activation regime, while at higher doping it saturates and slowly increases following the increase of disorder and formation of aggregates \cite{Desvignes2023}. In the parameter range of this work, we are close to $R_N$ saturation, and only little variations are observed when modifying $C_B$ (see Fig. \ref{fig:RT_B}). \\
The evolution of $T_c$ with $C_B$ is instead marked. $T_c$ initially increases roughly linearly with $C_B$, to attain a maximum at $C_B=8.9\, at.\%$, and then decreases, more slowly, in the saturation regime (Fig.\ref{fig:Tc_B}-a). The similar evolution of $T_c (C_B)$ and $n_B (C_B)$ suggests that $T_c$ is controlled by the active concentration $n_B$. However, we observe that $T_c$ keeps increasing with $C_B$ even \textit{after} the saturation of the hole concentration $n_B$ at $C_B=7.8 \,at.\%$. 
Thus, the question arises if superconductivity in silicon, besides being controlled by the carrier density, might be tuned through the strain induced, at the same time, by the smaller B atoms. The demonstration of superconductivity in SiGe by ultra-doping with B, in addition to its intrinsic interest associated to the role played by SiGe in classical (and quantum) electronics, opens the way to an experimental answer to this question. The independent incorporation of B and Ge allows indeed to address, independently, the role of the carrier concentration and the strain for superconductivity.\\
\begin{figure}[b]
		\centering
		\includegraphics[width=\columnwidth]{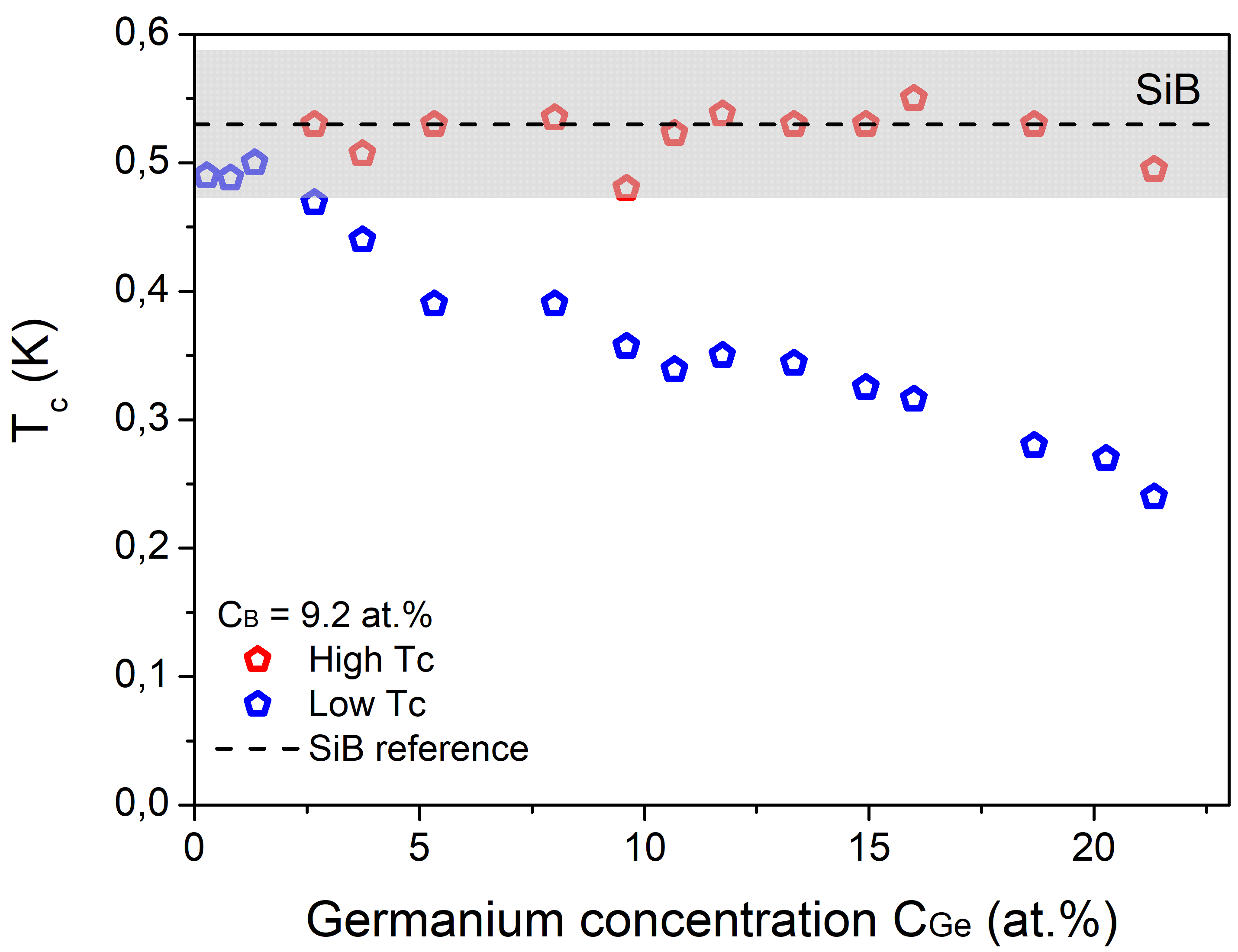}
		\caption{\label{fig:Tc_Ge} Superconducting critical temperatures of the two resistive transitions $T_{c,l}$ (blue) and $T_{c,h}$ (red) observed in $SiGeB$ series GILD-Ge (see Fig.\ref{fig:RT_B}), for a fixed $C_B = 9.2 \, at.\%$ and as a function of Ge concentration $C_{Ge}$. The dotted line corresponds to the $T_c$ of the reference $SiB$ sample with $C_B = 9.2 \,at.\%$ and no Ge, realized in the same run as the GILD-Ge series. The gray line is a guide to the eye.
  }
	\end{figure}
\subsection{Superconductivity evolution with Ge and B in $Si_{1-x}Ge_{x}:B$}
Fig. \ref{fig:RT_B}-b shows the $R(T)$ superconducting transitions of typical $SiGeB$ layers from the GILD-Ge sample series, with constant $C_B = 9.2\,$at.$\%$ and varying Ge concentration $C_{Ge}=0.8 - 21.3\,$at.$\%$. While at small Ge content the transitions are nearly on top of the reference $SiB$, single, transition, for higher $C_{Ge}$ the curves show two transitions. The first transition, $T_{c,h}$, at higher temperature, accounts for $23-24\,\%$ of the total resistance. The second transition, accounting for $\sim 75\%$ of the total resistance, is similarly sharp, with $\Delta T \sim 0.05\,$K$\sim 13 \%$, and is characterized by a lower transition temperature $T_{c,l}$. The decrease of $T_{c,l}$ with Ge concentration observed in Fig. \ref{fig:RT_B}-b is a general occurrence for all B concentrations, and not particular to the fixed $C_B$ of the curves displayed. Indeed, as shown in Fig. \ref{fig:Tc_B}, $T_{c,l} (C_B)$ follows for each $SiGeB$ series a dependence that mimics that of $SiB$, but shifted towards lower $T_c$ values, with a shift that increases with $C_{Ge}$. A strong disorder induced by the Ge incorporation might explain such $T_{c,l}$ reduction. However, an important disorder would be evident in the normal state square resistance at low temperature, $R_{N,sq} \sim 3\,\Omega$, which instead remains well below the resistance quantum (see Fig. \ref{fig:RT_B}). Moreover, $R_N$ is only little affected by the Ge incorporation, and a slight $R_N$ reduction with $C_{Ge}$ is even observed, probably related to a better carrier mobility in SiGe despite the scattering induced by Ge random position in the lattice. Thus, the large $T_c$ suppression ($\delta T_{c,l}/T_{c,l} \sim 50\,\%$) cannot be explained by the low disorder ($\delta R_N/R_N \sim 10\,\%$).\\
Fig. \ref{fig:Tc_Ge} shows the evolution of $T_{c,l}$ and $T_{c,h}$ as a function of the Ge concentration. $T_{c,h}$ is on average constant, globally independent of the Ge content, and its value is consistent within $7\%$ with the $T_c$ expected, in the absence of Ge, for a $SiB$ layer of the same B doping. In addition, the measured critical magnetic field $H_{c2,h}$, is also compatible with the $SiB$ reference, $H_{c2}\sim 200$ to $1000\,$G in the $C_B$ range examined. Fitting $H_{c2,h} (T)$ (see Methods), it is possible to extract the superconducting coherence length $\xi_{SiGeB,h}$. For $C_B = 9.2\,at.\%$, we find $\xi_{SiGeB,h}=59-65\,$nm, similar to $SiB$ $\xi_{SiB}=60\,$nm but with $\sim 10\,\%$ larger value associated to a higher diffusion coefficient (as also observed in $R_N$).\\
In contrast, we find, for the low temperature transition $T_{c,l}$, a strong dependence with $C_{Ge}$, a suppressed $H_{c2,l} \sim 150$ to 500 G, and a larger $\xi=70$ to 150$\,$nm, the result of a doubled diffusion coefficient as compared to $\xi_{SiB}=50$ to 100$\,$nm.  \\
Thus, while at low Ge content we observe the behavior of a homogeneous $SiGeB$ layer, two phases are present at high Ge content: one depleted in Ge, behaving as pure $SiB$ with only a slightly increased disorder as a result of the Ge incorporation processes, the other deeply affected by the incorporated Ge, with a doubled diffusion coefficient and a suppressed $T_c$.\\
\subsection{Superconductivity evolution with lattice deformation}
\begin{figure}[t]
		\centering
		\includegraphics[width=\columnwidth]{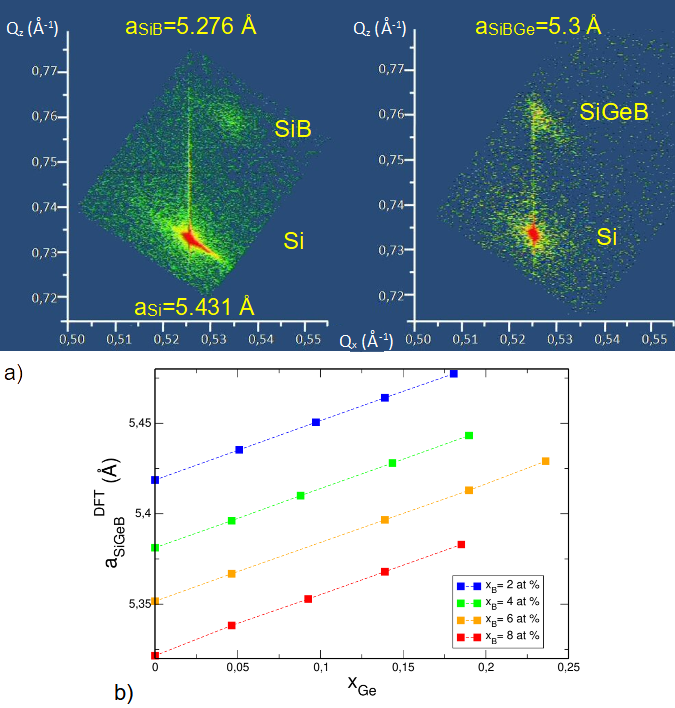}
		\caption{\label{fig:maps} a) XRD reciprocal maps along (224) direction for (left) a $SiB$ sample with $C_B=9.2\,at.\%$ and (right) a $SiGeB$ sample with $C_B=9.2\,at.\%$ and  $C_{Ge}=10.7\,at.\%$. Intensity is depicted in color contrast, with cold colors for the lower signal and hot colors for higher intensities. $Q_x$ and $Q_z$ correspond to the in-plane and out-of-plane wavevectors. The lattice parameter extracted from the XRD measurements is noted on each image. b) Optimized lattice constant for the ternary $SiGeB$ bulk alloy calculated within Density Functional Theory (DFT) in the generalized gradient approximation (GGA) as a function of the Ge fraction, $x_{Ge}$ and for different values of the B fraction, $x_B$. The $x_{Ge}$ value lies in the range of 0 to $25\,at.\%$ while $x_B$ varies between 2 and $8\,at.\%$.
  }
\end{figure}

In order to understand the role of Ge on $SiGeB$ superconductivity, and as Ge concentration does not directly affect the carrier concentration, we focus on the structural properties of the layer. In particular, we examine the deformation induced by both the Ge-induced compressive strain and the opposite B-induced tensile strain.
X-Ray Diffraction maps around the (224) reflection are realised to image both the in-plane and out-of-plane layer deformations (see Methods). 
Two samples are shown in Fig. \ref{fig:maps}: a $SiB$ layer, with $C_B = 9.2\,at.\%$, and a $SiGeB$ layer, with the same B concentration and $C_{Ge}=10.7\,at.\%$. The $SiB$ layer is partially relaxed, with an in-plane lattice constant smaller than that of the Si substrate, as visible from the larger $Q_x$ wavevector in $SiB$ as compared to Si. However, upon incorporation of Ge, the layer evolves back to a nearly fully strained configuration, with only the beginning of strain relaxation. This is the result of Ge partially compensating the B induced strain, as $a_{Ge} = 5.6578\,\AA > a_{Si} = 5.4307\,~\AA > a_{B} = 3.74\,~\AA$ \cite{Bisognin2007}.  From the in-plane and out-of-plane wavevectors $Q_x$ and $Q_z$, we extract the lattice parameter of the $SiB$ ($SiGeB$) layers, reported in Fig. \ref{fig:maps}-a, with $ (Q_{Si} - Q_{SiB})/Q_{SiB} = (a_{SiB} - a_{Si})/a_{Si}$ and $Q = \sqrt{Q_x^2 + Q_z^2}$.

\begin{figure}[t]
		\centering
		\includegraphics[width=\columnwidth]{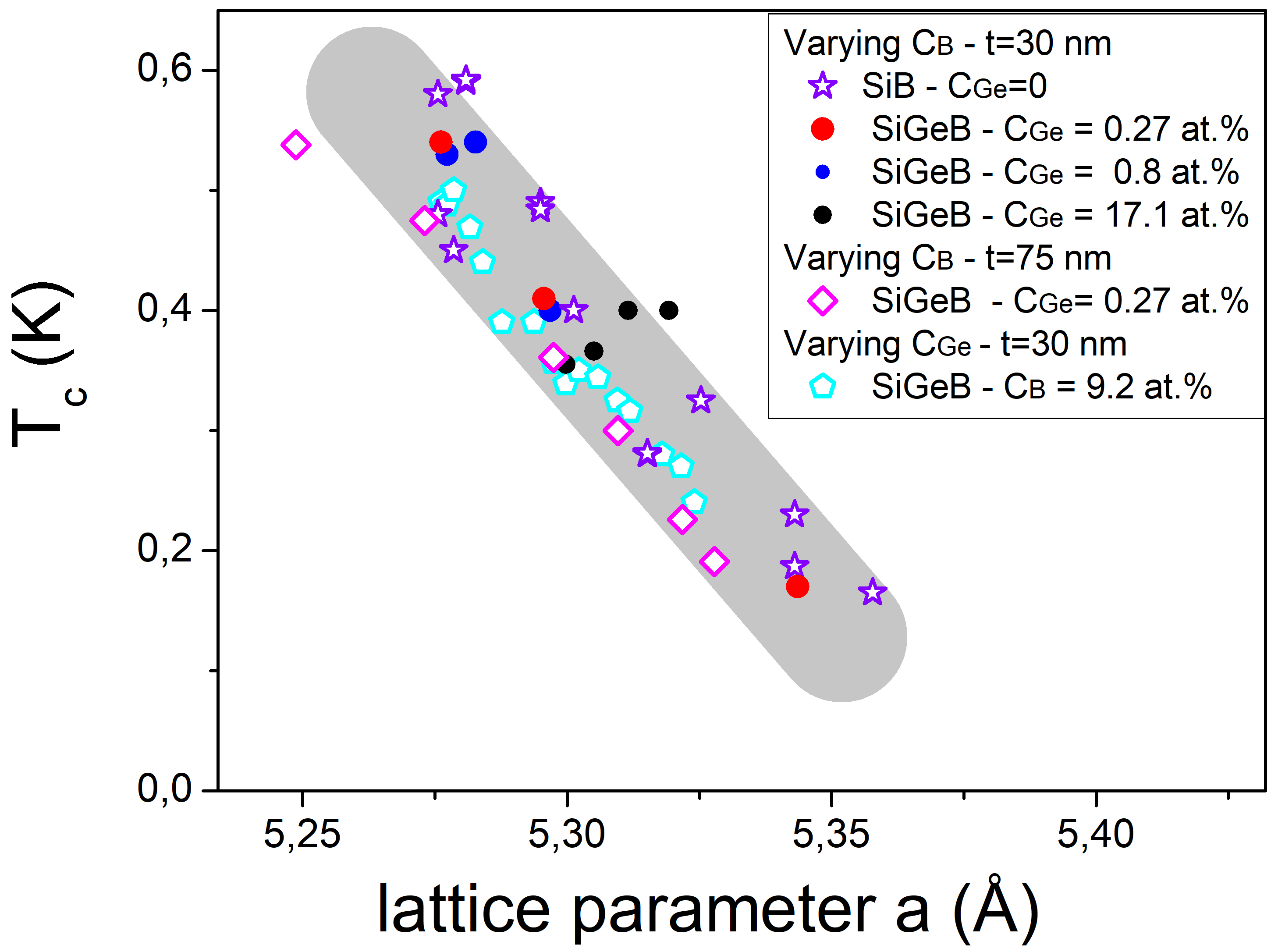}
		\caption{\label{fig:Tc_a} Superconducting critical temperature vs. lattice parameter calculated with Eq.\ref{eq_SiGeV_DFT} for $SiB$ reference series, GILD-5, GILD-15, GILD-200, GILD-Ge and Ge implanted (see Table \ref{Table} All series are plotted, for $C_B \leq 9.2\,at.\%$, to avoid the region where B aggregates are present, affecting the estimation of the lattice parameter.
  }
\end{figure}

Having established the experimental lattice parameters available from the limited number of XRD maps, we explore their dependence on B and Ge concentration by performing Density Functional Theory (DFT) calculations in the generalized gradient approximation (GGA) for the ternary $SiGeB$ bulk alloy. Special Quasi Random Structure approach~\cite{Zunger1990} is used to extract from the ensemble of all the possible random configurations only those which provide the most accurate approximation to the true random alloys. Further details of the DFT simulations are presented in the Methods section. For the pure SiGe alloy we find that, in the low Ge concentration regime, the behavior of the lattice constant follows the Vegard's law for binary semiconductors~\cite{Denton1991,Vegard1921}. Indeed, the DFT calculated lattice parameter of the alloy linearly increases with Ge fraction $x_{Ge}$, according to $a_{SiGe} = a_{Si} \cdot x_{Si} + a_{Ge} \cdot (1-x_{Si})$, where a$_{Si}$= 5.449~\AA~and a$_{Ge}$= 5.789~\AA~are the DFT-GGA lattice parameter of pure Si and pure Ge, respectively. Once the pure SiGe alloy case analyzed, we calculate the dependence of the lattice constant on the B concentration for the ternary $SiGeB$ bulk alloy. We consider a B fraction, $x_{B}$, ranging from 2$\,at.\%$ to 8$\,at.\%$ and we vary the Ge composition from 0 to 25$\,at.\%$, matching the experimental parameter range. As is shown in Fig.~\ref{fig:maps}-b, increasing $x_{B}$ lowers the value of a$_{SiGe}$ while maintaining the linear Vegard's behavior. These results demonstrate that, in this chemical composition regime, the use of a linear interpolation of the three alloy components is theoretically justified and can be summarized in the following equation: 
\begin{equation}
   a_{SiGeB} = a_B \, \cdot x_B + a_{Ge} \, \cdot x_{Ge} + a_{Si} \, \cdot (1-x_B - x_{Ge})
\label{eq_SiGeV_DFT}
\end{equation}
where $a_{B}$ is determined through a constrained fit (with a$_{Si}$ and a$_{Ge}$ fixed to their GGA values) to be 3.81~\AA, which is in very good agreement with the experimental value measured in Ref.~\cite{Bisognin2007}. Even though, due to the well-known GGA underbinding tendency~\cite{Zhang2018}, the simulated pure elements lattice parameters are slightly overestimated if compared with the experimental values (by a few percent difference), the theoretical results fully validate Eq.\ref{eq_SiGeV_DFT}.
We thus employ Vegard's law (Eq.~\ref{eq_SiGeV_DFT}) as a function of $x_{B}$ and $x_{Ge}$ to predict the lattice parameters of the samples shown in Fig. \ref{fig:maps}, by taking as input the experimental values of $C_{B}$ and $C_{Ge}$ and the experimentally determined  $a_{Si}=5.4307$~\AA, $a_{Ge}=5.6578$~\AA~ and $a_B= 3.74$~\AA, the B and Ge fractions $x_j = C_j/n_{Si}$ being calculated in respect to the pure Si density $n_{Si}  =5\times 10^{22} \, cm^{-3}$. For the samples analyzed in Fig. \ref{fig:maps}, we obtain $a_{SiB} = 5.2755$~\AA (vs. $a_{SiB, XRD} = 5.276\,\pm 0.005$~\AA), and $a_{SiGeB} = 5.300$~\AA (vs. $a_{SiGeB, XRD} = 5.299\,\pm 0.005$~\AA), an excellent agreement, well within the error associated to 'pointing' uncertainty on the XRD maps.\\
With both experimental and numerical validation, we extend the 3-elements Vegard's law to calculate the lattice parameters $a$ for the samples for which no XRD is available. To correctly estimate $a_{SiB}$ and $a_{SiGeB}$, only the substitutional dopant concentration providing a lattice deformation is relevant. We thus exclude from the following analysis the samples with $C_B > 9.2\,at.\%$, the concentration range where aggregates appear, rendering inaccurate the estimation of the substitutional concentration \cite{Hallais2023}. The concentration range between the fully activate regime and the saturation regime $6\,at.\% < C_B \leq 9.2\,at.\%$ is however included, as the still-high activation (ratio of active to total B concentration $>75\,\%$) is limited in this region by substitutional inactive B complexes formed by a few atoms (B dimers, trimers) \cite{Desvignes2023}. Such complexes also induce a lattice deformation, whose value differs however from that of isolated B atoms \cite{Bisognin2006, Bisognin2006-2}. We thus might expect a maximum error of $\sim 20\,\%$ on the deformation estimation for these complexes, but, as they account at most for $25\,\%$ of $C_B$ (and only at the highest concentrations), the final induced error on the lattice parameter is expected within a few $\%$.
In the case of Ge, the whole $C_{Ge}$ range is considered, as for the concentrations investigated here, Ge is expected to be fully substitutional \cite{Fossard2008}. 
\\
The dependence of $T_c$ on the lattice parameter $a$ calculated from $C_B$ and $C_{Ge}$ is shown in Fig. \ref{fig:Tc_a}. It is remarkable that all sample series collapse in a common linear trend: $SiB$ samples see their lattice parameter decrease with B doping and 'move' towards higher $T_c$ from right to left; $SiGeB$ samples with a fixed $C_B$ see the lattice parameter increase due to the Ge incorporation, and move from left to right to lower $T_c$, over $SiB$ samples with smaller B concentrations. The multiple methods employed to incorporate the Ge do not seem to affect significantly the global result, and neither does the difference between the 30 nm and 80 nm thick samples series. 
The series to series deviations from the average $T_c (a)$ observed in Fig. \ref{fig:Tc_a} are associated to uncertainties in the lattice parameter of $\delta a/a \sim 0.7\,\%$, and can be traced back to the uncertainty in the determination of the deformation associated to the few-atoms complexes.
It is noteworthy that modifying the lattice parameter by $\delta a /a = - 1\,\%$ leads to a large change in the superconducting critical temperature of $\delta T_c /T_c = 50\,\%$ .
A strong dependence of $T_c$ with the lattice parameter is reported for other superconductors, such as $In_xTe$ \cite{Kriener2022} or covalent superconductors (like Si and SiGe), such as superconducting B doped diamond ($\delta T_c /T_c \sim 64\,\%$ for $\delta a /a \sim 0.2\,\%$) \cite{Zhou1992} or $K_3C_{60}$ and $Rb_3C_{60}$ fullerenes ($\delta T_c /T_c \sim 83\,\%$ for $\delta a /a \sim 4\,\%$) \cite{Brazhkin2006}. \\
Such strong increase of $T_c$ upon reduction of the lattice parameter can be associated to the softening of phonon modes, which increases the electron-phonon coupling $\lambda$ through the increase of the electron-phonon potential $V_{e-ph}$ \cite{Boeri2004}. The incorporation of $Ge$ in the lattice by nanosecond laser annealing allows thus addressing specifically the electron-phonon potential, through a finely tuned lattice parameter.

\section{Conclusions}
In conclusion, we demonstrate superconductivity in $Si_{1-x}Ge_{x}:B$ epilayers by nanosecond laser ultra-doping with B. The B concentration reached, well above the solubility limit thanks to this out-of-equilibrium technique, is $C_B=1.5$ to $11\,at.\%$, with 100\% activation rate below $C_B=6\,at.\%$ and over 75\% up to $C_B=9.2\,at.\%$. The Ge fraction explored is $x=$0 to 0.21.  Ge is incorporated in $SiB$ in three different ways: 1) through a precursor gas by Gas Immersion Laser Doping; 2) by ion implantation, followed by nanosecond laser annealing; 3) by UHV-CVD growth of a thin Ge layer, followed by nanosecond laser annealing. The 30 nm and 80 nm thick $Si_{1-x}Ge_{x}:B$ epilayers display a zero resistance state, with superconducting critical temperature $T_c$ varying with B and Ge concentration from 0 to 0.6 K, a superconducting critical field $H_{c2} = 150$ to 500 G, and a superconducting coherence length $\xi=70$ to 140$\,$nm, larger than $SiB$ layers of equal concentration due to a doubled diffusion coefficient associated to Ge incorporation. To understand the $T_c$ evolution, we turn towards BCS theory, which predicts a $T_c$ exponential evolution with the electron-phonon coupling constant $\lambda = N(E_F) V_{e-ph}$, the product of the density of state at Fermi level and the electron-phonon interaction potential. Starting with the simplest $SiB$ alloy, we observe an initial increase of $T_c$ with $C_B$, that can be associated to the increase of the charge carrier density  ($n_B = 1.5 - 7 \,at.\%$), and as a consequence of $N(E_F)$. However, $T_c$ keeps increasing even when the hole concentration $n_B$ saturates at $C_B=7.8\,at.\%$, following the formation of B aggregates. We thus explore the role played by the structural deformations on superconductivity, by fine-tuning the strain through the modulation of the Ge concentration at fixed carrier density. To estimate the lattice parameter modulation with B and Ge, we validate Vegard's law for the ternary $SiGeB$ bulk alloy by DFT-GGA calculations. The theory is in excellent agreement with X-Ray Diffraction maps, allowing to measure both the in-plane and the out-of-plane lattice deformation.  By correlating the $T_c$ with the calculated lattice parameter, we observe a global linear dependence, common for both $Si:B$ and $Si_{1-x}Ge_{x}:B$ layers, and independent on the Ge incorporation method or on the sample thickness, with $\delta T_c/T_c \sim 50\,\%$ for $\delta a/a \sim 1\,\%$, thus highlighting the importance of structural strain, at fixed carrier concentration.

\subsection*{Acknowledgements}
We are grateful for support from the French CNRS RENATECH network, the Physical Measurements Platform of University Paris-Saclay, the French National Research Agency (ANR) under Contract No. ANR-16-CE24-0016-01, ANR-19-CE47-0010-03 and ANR-22-QUA2-0002-02. M.A. and M. T. acknowledge the ANR AMPHORE project (ANR-21-CE09-0007). 

\subsection*{Methods}
\subsubsection{Gas Immersion Laser Doping}
GILD is performed in an ultra high vacuum (UHV) reactor of base pressure $10^{-9}-10^{-10}$ mbar, to ensure a minimal impurity incorporation during the melt phase. A puff of the precursor gas ($BCl_3$ or $GeCl_4$) is injected onto the substrate surface, saturating the chemisorption sites ($p\sim 10^{-5} mbar$). A pulse of excimer XeCl laser ($\lambda$ = 308 nm, pulse duration 25 ns, working frequency 2 Hz) melts the substrate, the light being completely and instantly converted into thermal energy in the top 7 nm. During the melted phase, the chemisorbed atoms diffuse in the liquid. At the end of the laser pulse, an epitaxial out-of-equilibrium recrystallization takes place from the substrate at a speed of $\sim 4\,$m/s \cite{Wood1984}, achieving concentrations larger than the solubility limit. When the crystallization front reaches the surface, the excess impurities contained in the liquid are expelled outwards, such as Cl whose segregation coefficient is close to 0 \cite{Bourguignon1995}. Thanks to a careful optical treatment of the laser beam, the energy density at the 2mmx2mm sample level has $\sim$1.2$\%$ spatial homogeneity. This ensures the homogeneity of the layer thickness, resulting in a flat, straight, and sharp (a few nm thick) interface of the SiGe with the substrate. Since the laser absorption is sensitive to the layer doping level, a fixed laser energy results in an increasing layer depth. In order to obtain a constant doped depth independent of the B or Ge content, we measure the time-resolved reflectometry, and maintain a constant melt time during the doping by decreasing progressively the laser energy.\\
The B doping is always performed before the Ge incorporation. Indeed, a homogeneous distribution is expected even when the B is further submitted to the subsequent process time of the Ge incorporation.  In contrast, the Ge profile is expected to evolve toward the surface, depleting the bottom of the layer \cite{Fossard2008}. To keep the Ge profile the most homogeneous possible, we incorporate Ge last, to minimise the time spent by Ge atoms in the liquid phase.\\

\subsubsection{UHV-CVD growth of Ge/Si}
The epitaxial growth of Ge on Si is carried out in an UHV-CVD system  with a base pressure of $10^{-10}$ mbar. Pure $SiH_4$ and $GeH_4$ diluted at 10$\%$ in $H_2$ are used as gas sources. After a modified Shiraki chemical cleaning \cite{Nguyen2003} the substrates are slowly annealed up to 700$^{\circ}$C, the pressure being maintained below $7 \cdot 10^{-9}$ mbar. Afterwards, the chemical surface oxide is removed by flashing at 990$^{\circ}$C, maintaining the low pressure. After the deposition of a Si buffer layer at 700$^{\circ}$C under a pressure of $4 \cdot 10^{-4}$ mbar, the Ge heteroepitaxy at 330$^{\circ}$C is initiated at a total pressure of $7 \cdot 10^{-3}$ mbar. The growth time is settled at 20 min in order to achieve 6 nm of Ge \cite{Halbwax2005,Hallais2023a}.\\
\subsubsection{Hall measurements}
The transverse voltage $V_H$ is measured in a magnetic field perpendicular to the layer, at room temperature, with $V_H /I= \gamma \frac{B}{e n_B t}$, $\gamma = 0.68$ the Hall mobility factor \cite{Lin1981}, $I$ the bias current (10$\mu$A), $B$ the applied magnetic field (0 to 2 T), $e$ the electron charge and $t$ the layer thickness.\\

\subsubsection{Measurement of the critical magnetic field $H_{c2} (T)$}
$R(T)$ superconducting transitions are measured for fixed values of a perpendicular magnetic field from 0 to 55 mT. Both the $SiB$ reference samples series and a few selected $SiGeB$ samples are studied [$(C_B, C_{Ge}) = (9.2, 8) \,at.\%$; $(9.2, 21.3) \,at.\%$; $(8, 10.7) \,at.\%$; $(10.6, 10.7) \,at.\%$].
In the temperature range $T=0.2 - 0.5\,$K, $SiB$ follows the expected trend for a thin superconducting film near $T_c$: $\mu_0 H_{c2} = \frac{3\Phi_0}{(2\pi^2 \,\xi^2)}(1-\frac{T}{T_c})$. The extracted coherence length is $\xi=60\,$nm, in agreement with previous measurements \cite{Bonnet2019}. $SiGeB$ $T_{c,h}$, measured in the same temperature range, follows the same law, with $\xi=59-65\,$nm. These values, in agreement with $SiB$ results, confirm the observed absence of Ge. The slightly larger values might be a result of a small additional disorder as the layers have been submitted to supplementary processes to incorporate Ge. Indeed, the diffusion coefficient $D$ affects $\xi$, as $\xi = \sqrt{(\hbar D/1.76 k_B T_c)}$.\\

\subsubsection{X-Ray Diffraction} 
The diffractograms are realized with a Rigaku Smartlab XRD system with Cu-K$\alpha$1 radiation of wavelength 1.54056$\,\r{A}$, operated at 45 kV and 200 mA. The x-ray beam is narrowed to measure only the central, homogeneous part of the laser annealed spot.  To avoid the contribution of the gold contacts in the diffractogram, the contacts were removed by a KI Au-etch followed by 1 min dip in a 10$\%$ HF solution to remove Ti.

\subsubsection{DFT simulations}
DFT calculations were performed by using the SIESTA package~\cite{Soler2002} whose numerical atomic orbitals basis set allows treating large systems with an affordable computational cost. The exchange–correlation energy functional was approximated using the generalized gradient approximation (GGA) as implemented by Perdew, Burke, and Ernzerhof (PBE functional)~\cite{Perdew1996}. Only valence electrons have been taken into account with core electrons being replaced by norm-conserving pseudopotentials of Troullier-Martins type. An optimized double-$\zeta$ polarized basis set was used for Si while a double-zeta plus two polarization orbitals (DZP2) basis set was chosen for both Ge and B. All the equilibrium ground state unit cells and geometries were obtained from conjugate-gradients structural relaxation using DFT forces through the Hellman-Feynman theorem. The structures were relaxed until the force on each atom was smaller than 0.01 eV/~\AA. The cutoff of the grid used for the real space integration was set to 300 Ry while the self-consistent cycle tolerance for solving the Kohn-Sham equations was set to 10$^{-4}$ eV. A uniform Monkhorst-Pack grid with 3$\times$3$\times$3 k-points was employed to sample the Brillouin zone. A supercell of 216 atoms was considered, which corresponds to a 3$\times$3$\times$3 supercell of the conventional 8-atom cell of bulk Si. To take into account the random nature of the alloy the Special Quasi Random Structure (SQS) approach~\cite{Zunger1990} was adopted. The SQS configurations were generated using the ATAT code~\cite{Van2002} and considering that each atom in the supercell can be replaced with a probability depending on its concentration, as shown in Ref.~\cite{Van2013}. 
As a starting point, a pure Si supercell was considered and the Ge fraction of atoms was varied from 0 to 0.25 for several SQS configurations. For each of these configurations, the lattice parameter of the alloy was averaged over the different configurations and over the three cubic crystal axis to minimize the error due to numerical fluctuations during optimization. Once the pure SiGe alloys case was treated, the dependence of the lattice parameter on the B concentration was calculated. B concentrations from 2\% to 8\% and Ge composition from 0 to 25\% were considered. All the studied systems are substitutional solid alloys in which Si, Ge, and B can occupy only substitutional lattice sites (interstitial are not taken into account).

\bibliographystyle{unsrt}
\bibliography{biblio}

\end{document}